\documentclass[aip, reprint, superscriptaddress, showpacs]{revtex4-1}
\usepackage{graphicx}
\usepackage{natbib, verbatim, ulem}
\usepackage{color, xcolor}

\newcommand{\vdW}{vdW$^{\rm surf}$}

\begin{document}

\title{The role of the van der Waals interactions in the adsorption of anthracene and pentacene on the Ag(111) surface}
\author{Juliana M. Morbec}
\email{jmmorbec@gmail.com}
\author{Peter Kratzer}
\affiliation{Faculty of Physics, University of Duisburg-Essen, Lotharstrasse 1, 47057 Duisburg, Germany}

\begin{abstract}
Using first-principles calculations based on density-functional theory (DFT) we investigated the effects of the van der Waals (vdW) interactions 
on the structural and electronic properties of anthracene and pentacene adsorbed on the Ag(111) surface. 
We found that the inclusion of vdW corrections strongly affects the binding of both anthracene/Ag(111) and pentacene/Ag(111), 
yielding adsorption heights and energies more consistent with the experimental results than 
standard DFT calculations with generalized gradient approximation (GGA). 
For anthracene/Ag(111) the effect of the vdW interactions is even more dramatic: we found that ``pure'' 
DFT-GGA calculations (without including vdW corrections) result in preference for a tilted configuration, 
in contrast to experimental observations of flat-lying adsorption; including vdW corrections, on the other hand, alters the binding geometry of anthracene/Ag(111), 
favoring the flat configuration. 
The electronic structure obtained using a self-consistent vdW scheme was found to be nearly indistinguishable from the conventional DFT electronic structure once the correct vdW geometry is employed for these 
physisorbed systems. Moreover, we show that a vdW correction scheme based on a hybrid functional DFT calculation (HSE) results in an improved description of the highest occupied molecular level of the adsorbed molecules. 
\end{abstract}

\maketitle

\section{INTRODUCTION} 
Interfaces between organic molecules and metal surfaces are fundamental constituents of numerous organic optoelectronic devices 
and play important roles in their performance. 
Thus, understanding the adsorption of organic molecules on metal surfaces 
and the interactions at such organic/metal interfaces is crucial to develop efficient devices. 

Anthracene (C$_{14}$H$_{10}$) and pentacene (C$_{22}$H$_{14}$) 
adsorbed on the Ag(111) surface are particularly interesting systems because of the frequent use of silver as electrode  
and the promising properties 
of the anthracene and pentacene compounds; both compounds exhibit 
high carrier mobility \cite{Nelson-APL72.1854, *Jurchescu-APL84.3061} 
and excellent electroluminescence and photoluminescence \cite{Tao-ChemPhysLett429.622}, 
and have great potential for use in organic light-emitting diodes \cite{Kitamura-APL83.3410, *Dobbertin-APL82.284} and 
organic field effect transistors \cite{Butko-APL83.4773, *Dimitrakopoulos-Science283.822}. 
Experimental studies have reported that both anthracene and pentacene monolayers weakly adsorb on Ag(111), 
as characteristic of physisorption or weak chemisorption 
\cite{Yannoulis-SurfSci241.325, Shimooka-Langmuir-2001, Frank-JCP89.7569, Koch-OrgElectronics7.537, Dougherty-JPhysChemC112.20334, Duhm-ApplMatInt5.9377}, 
which suggests that van der Waals interactions (vdW) play an important role in the binding of these systems. 

Van der Waals interactions are weak interactions that 
originate from non-local correlations between electronic charge fluctuations. 
They are the dominant interactions in physisorbed and weakly chemisorbed systems 
and are known to strongly affect the 
stability and structure of organic/metal interfaces \cite{Tkatchenko-MRS35.435, Li-PhysRevB.85.121409, Liu-AccChemRes47.3369, Liu-PhysRevB.86.245405}. 
Therefore, various schemes have been devised how to include vdW interactions 
into density functional theory (DFT). While some are formulated directly as functionals of the 
electron density \cite{Dion-PhysRevLett.92.246401, Langreth-JPCM21.084203}, the more common and convenient schemes simply add a correction 
to the DFT total energy in the form of a pairwise interaction potential \cite{Grimme-JCompChem27.1787,PhysRevLett.102.073005}. 

In this work, 
we present a first-principles study on 
the effects of the vdW interactions on the structural and electronic 
properties of anthracene and pentacene adsorbed on the Ag(111) surface. 
We used both the vdW$^{\rm surf}$ approach \cite{PhysRevLett.108.146103} 
and the many-body dispersion (MBD) method \cite{PhysRevLett.108.236402, Ambrosetti-JChemPhys140.18A508} to 
treat the vdW interactions in our calculations. 
While the first works with an additive pairwise potential, albeit with 
density-dependent coefficient, the second approach is able to take cooperative 
effects into account. 
Both the \vdW \ and MBD methods have been shown to provide reliable adsorption energy and height for interfaces between organic molecules (such as benzene and PTCDA) 
and coinage metal surfaces~\cite{Maurer-JCP143.102808, Ruiz-PhysRevB.93.035118, Carrasco-JCP140.084704}. 
Our results show that the inclusion of vdW corrections is crucial to correctly describe the flat adsorption geometry  
of anthracene on Ag(111); calculations using ``pure'' generalized gradient approximation~(GGA), without including vdW corrections, resulted in a tilted configuration for 
anthracene/Ag(111), while the use of the vdW$^{\rm surf}$ approach yielded flat molecular orientation, 
consistent with experimental observations. 
We also found that the adsorption heights and the adsorption energies 
in both anthracene/Ag(111) and pentacene/Ag(111) are strongly affected by the vdW treatment. 
On the other hand, we noticed that including the density-dependence inherent 
in the \vdW approach has only tiny 
effects on the electronic structure of the systems. 
Lastly, we computed the change in the work function of the Ag(111) surface upon the adsorption of anthracene and pentacene; 
we found that anthracene 
induces a larger reduction in the work function than pentacene, 
consistent with the stronger physisorption character observed in anthracene/Ag(111) in comparison with pentacene/Ag(111). 

The rest of the paper is organized as follows: in Sec.~\ref{comp-details} we
describe the details of our first-principles calculations; in Sec.~\ref{results} we present and discuss our results
for the  structural and electronic properties of anthracene/Ag(111) and pentacene/Ag(111); and in Sec.~\ref{conclusions} 
we provide our conclusions.

\section{COMPUTATIONAL DETAILS} \label{comp-details}

All calculations were performed in the framework of the density-functional theory \cite{PhysRev.140.A1133} 
as implemented in the FHI-aims \cite{Blum20092175},  
an all-electron code that uses numeric atom-centered orbitals as basis functions. 
We used the generalized gradient approximation~(GGA) proposed by Perdew, Burke,
and Ernzerhof~(PBE) \cite{PhysRevLett.77.3865} for the exchange-correlation functional, as well as, the hybrid 
Heyd-Scuseria-Ernzerhof (HSE06) \cite{Heyd-JChemPhys118.8207, *Heyd-JChemPhys124.219906} functional. 
``Tight'' settings from the FHI-aims code were used in all calculations, with ``tier2''/``tier3''/``tier2'' basis sets for Ag/C/H in the PBE calculations 
and ``tier1''/``tier2''/``tier2'' basis sets for Ag/C/H in the calculations with the HSE functional. 
Convergence criteria of $10^{-5}$ electrons/\AA$^3$ and $10^{-5}$~eV 
were applied for the charge density and the total energy, respectively. 

To treat the vdW interactions we employed two approaches coupled to PBE and HSE functionals: 
(i) the vdW$^{\rm surf}$ approach \cite{PhysRevLett.108.146103}, which includes the collective electronic response of the substrate  
in the determination of the vdW parameters ($C_6$ coefficients, polarizabilities and  vdW radii) 
by combining the pairwise Tkatchenko-Scheffler (TS) method \cite{PhysRevLett.102.073005} with 
the Lifshitz-Zaremba-Kohn theory \cite{Lifshitz-SovPhysJETP2.73, PhysRevB.13.2270} for the vdW interaction between an atom and a solid surface, 
and (ii) the many-body dispersion (MBD) method \cite{PhysRevLett.108.236402, Ambrosetti-JChemPhys140.18A508},   
in which the atomic response functions are represented by a set of 
quantum harmonic oscillators and the screened long-range many-body vdW energy is computed using the adiabatic connection 
fluctuation-dissipation theorem within the dipole approximation. 
In the calculations with the vdW$^{\rm surf}$ approach we used the screened vdW parameters computed in Ref.~\onlinecite{PhysRevLett.108.146103}. 
Throughout the paper we will refer to the calculations without including vdW corrections as ``PBE calculations'' (or ``HSE calculations'', when 
the hybrid HSE functional is used), and we will use ``PBE(HSE)+vdW$^{\rm surf}$'' and ``PBE(HSE)+MBD'' for the calculations 
with the vdW$^{\rm surf}$ and MBD approaches, respectively. 
To examine the effects of the vdW interactions on the electronic properties of anthracene/Ag(111) and pentacene/Ag(111), we 
used the self-consistent \vdW (sc-\vdW) method \cite{Ferri-PhysRevLett.114.176802}, 
in which the functional derivative (with respect to the density) of the 
pairwise TS vdW potential is added to the exchange-correlation 
potential to form the total effective Kohn-Sham potential. 

In order to build the supercells for our calculations, we first computed the lattice constant of bulk silver;   
using a $16\times 16\times 16$ Monkhorst-Pack $k$-point mesh, we found $a=4.140$~\AA \ and $a=4.020$~\AA \ in the 
PBE and PBE+vdW$^{\rm surf}$ calculations, respectively, which are in close agreement with the experimental value  
of 4.09~\AA \ \cite{Moruzzi-PhysRevB.37.790} and with previous theoretical results \cite{Liu-NJP15.053046} obtained using PBE (4.149~\AA) and PBE+vdW$^{\rm surf}$ (4.007~\AA).  
We used the respective computed lattice constants in our PBE and PBE+vdW$^{\rm surf}$ calculations. 

Anthracene/Ag(111) was modeled using 
a $(2\sqrt{3}\times2\sqrt{3})R30^{\circ}$ surface unit cell 
(in accordance with the STM measurements of anthracene adlayers on Ag surfaces in perchloric acid solution 
reported by Shimooka et al. \cite{Shimooka-Langmuir-2001}) with a five-layer slab; 
for pentacene/Ag(111) we used a $(6\times3)$ surface unit cell (which is consistent with experimental measurements \cite{Dougherty-JPhysChemC112.20334})  
with a four-layer slab. A vacuum region of $\sim$30~\AA \ was used in both cases to avoid unphysical interaction between periodic images. 
In the geometry optimizations, the molecules and the top two silver layers were allowed to relax while the remaining bottom layers 
were constrained to their bulk positions; a force convergence criterion of $10^{-2}$~eV/\AA \ was applied for structural relaxations. 
We used $4\times 4\times 1$ and $3\times6\times1$ Monkhorst-Pack $k$-point meshes in the 
structure optimization and total-energy calculations of anthracene/Ag(111) and  pentacene/Ag(111) systems, respectively, 
whilst denser $k$-point sets, $12\times 12\times 1$ for anthracene/Ag(111) and 
$9\times18\times 1$ for pentacene/Ag(111), were employed to compute the density of states and the work functions. 
For the calculations of the work function we 
examined the convergence of our results with respect to the number of layers and the thickness of the vacuum region.

\section{RESULTS AND DISCUSSION}  \label{results}

\subsection{Anthracene on Ag(111)}

\begin{figure}[!b]
\centering
\includegraphics[scale=0.18]{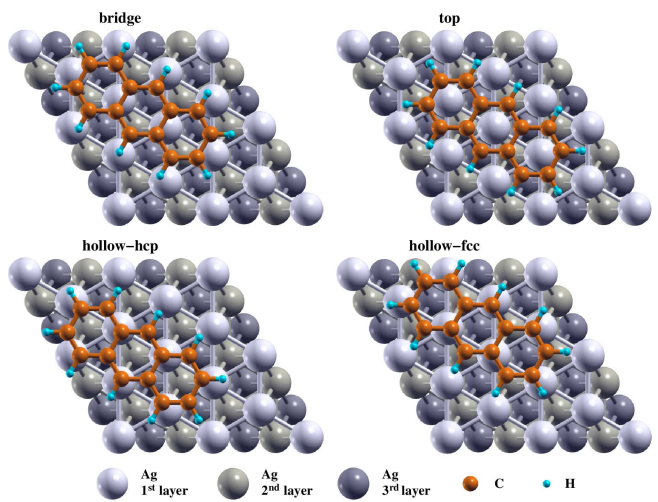}
\caption{\label{ANT-topview} Ball and stick representation of the adsorption sites (bridge, top, hollow-hcp, and hollow-fcc) 
examined for anthracene/Ag(111).}
\end{figure}

\begin{figure}[!b]
\centering
\includegraphics[scale=0.18]{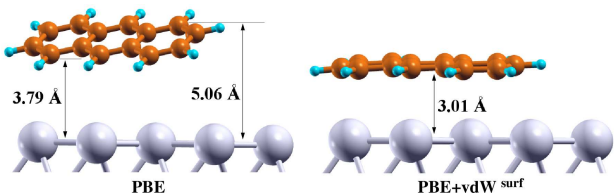}\\
\caption{\label{ANT-sideview} Equilibrium geometries of anthracene/Ag(111) (hollow-hcp configuration) obtained using the PBE (left)
and PBE+vdW$^{\rm surf}$ (right) methods.}
\end{figure}

We first determined the equilibrium geometry for anthracene/Ag(111). 
Considering only planar adsorption and 
taking as reference the central carbon ring, we examined four adsorption sites (see Fig.~\ref{ANT-topview}): 
bridge, where the central ring lies over a Ag-Ag bond; top, where the central ring is on top of a Ag atom from the top layer; 
hollow-hcp, where the central ring is on top of a 2nd-layer Ag atom; and hollow-fcc, where the central ring is on top of a 3rd-layer Ag atom. 
The most stable adsorption site was found to be hollow-hcp in both PBE and PBE+\vdW calculations 
(see Table~\ref{table-ANT-conf}), with total energy 1--4~meV smaller than the second most stable configuration, hollow-fcc. 
We found, however, that PBE calculations 
favored 
a tilted hcp-configuration (Fig.~\ref{ANT-sideview}) with the 
molecular short axis making a tilt angle 
of $\sim$$26.7^{\circ}$ with the substrate, 
in contrast to 
previous experimental observations \cite{Shimooka-Langmuir-2001, Yannoulis-SurfSci241.325, Frank-JCP89.7569} 
of flat adsorption of anthracene molecules on the Ag(111) surface. 
We found that the flat hollow-hcp configuration is 0.098~eV less stable than the tilted configuration, and it has adsorption height of 
4.095~\AA \ and adsorption energy of $-0.061$~eV. 
PBE+\vdW calculations, on the other hand, predicted flat geometry (see Fig.~\ref{ANT-sideview}) with 
adsorption energy of $-1.380$~eV (see Table~\ref{table-ANT-conf}), 
more than 1.3~eV stronger than that predicted by PBE calculations, and 
adsorption height of 3.015~\AA, significantly smaller than the adsorption height obtained without including vdW corrections (4.432~\AA) 
and in the range of typical molecule-substrate distances for aromatic molecules on metals
(2.8--3.2~\AA)~\cite{Hauschild-PRL.94.036106, Kilian-PRB.66.075412}. 
The adsorption height $d$ is calculated as the vertical distance between the molecule's center of mass 
and the 
average positions of the Ag atoms in the uppermost layer. 
In the hcp configuration, 
the surface-layer Ag atoms below the anthracene molecule relaxed 
inward by $\sim$$0.02$~\AA \ on average. 
The adsorption energy $E_{\rm ads}$ is computed using 
$$E_{\rm ads}=E_{\rm Mol/Ag(111)}-E_{\rm Ag(111)}-E_{\rm Mol},$$ 
where $E_{\rm Mol/Ag(111)}$, $E_{\rm Ag(111)}$ and $E_{\rm Mol}$ are the total energies of the adsorbed system, 
the bare Ag(111) surface, and the single molecule in a large supercell, respectively.

\begin{table}[!t]
\centering
\caption{\label{table-ANT-conf}Relative total energies $\Delta E_T$ (in eV), 
adsorption energies $E_{\rm ads}$ (in eV), and adsorption heights $d$ (in \AA)  
for anthracene/Ag(111) at different adsorption sites, computed using the PBE and PBE+\vdW methods. 
The equilibrium orientation of the molecule with respect to the substrate (flat or tilted) is indicated within brackets. 
$\Delta E_T$ is calculated as the difference in total energy of a given configuration and the most stable configuration (hcp). 
The adsorption height is determined as the vertical distance between the center of mass of the molecule 
and the 
average positions of the Ag atoms in the topmost layer. 
The values in parentheses in column 6 are the adsorption energies obtained with the PBE+MBD method at the PBE+\vdW geometry.}
\begin{ruledtabular}
\begin{tabular}{lcccccc}
     & \multicolumn{3}{c}{PBE} & \multicolumn{3}{c}{PBE+\vdW}\\
\cline{2-4}\cline{5-7}
site & $\Delta E_T$ & $E_{\rm ads}$ & $d$ & $\Delta E_T$  & $E_{\rm ads}$ & $d$\\
\hline
bridge & $0.106$ & $-0.060$ & $4.026$    & $0.074$ & $-1.313$   & $3.036$ \\
       &       &	& [flat]   &       & ($-0.856$) & [flat] \\
top    & $0.111$ & $-0.057$ & $4.112$    & $0.155$ & $-1.255$   & $3.047$ \\
       &       &        & [flat]   &       & ($-0.763$) & [flat] \\
hcp    & $0.000$ & $-0.065$ & $4.432$    & $0.000$ & $-1.380$   & $3.015$ \\
       &       &        & [tilted] &       & ($-0.925$) & [flat]\\
fcc    & $0.001$ & $-0.064$ & $4.440$    & $0.004$ & $-1.363$   & $3.017$ \\
       &       &        & [tilted] &       & ($-0.891$) & [flat]\\
\end{tabular}
\end{ruledtabular}
\end{table}

\begin{figure}[!t]
\centering
\includegraphics[scale=0.41]{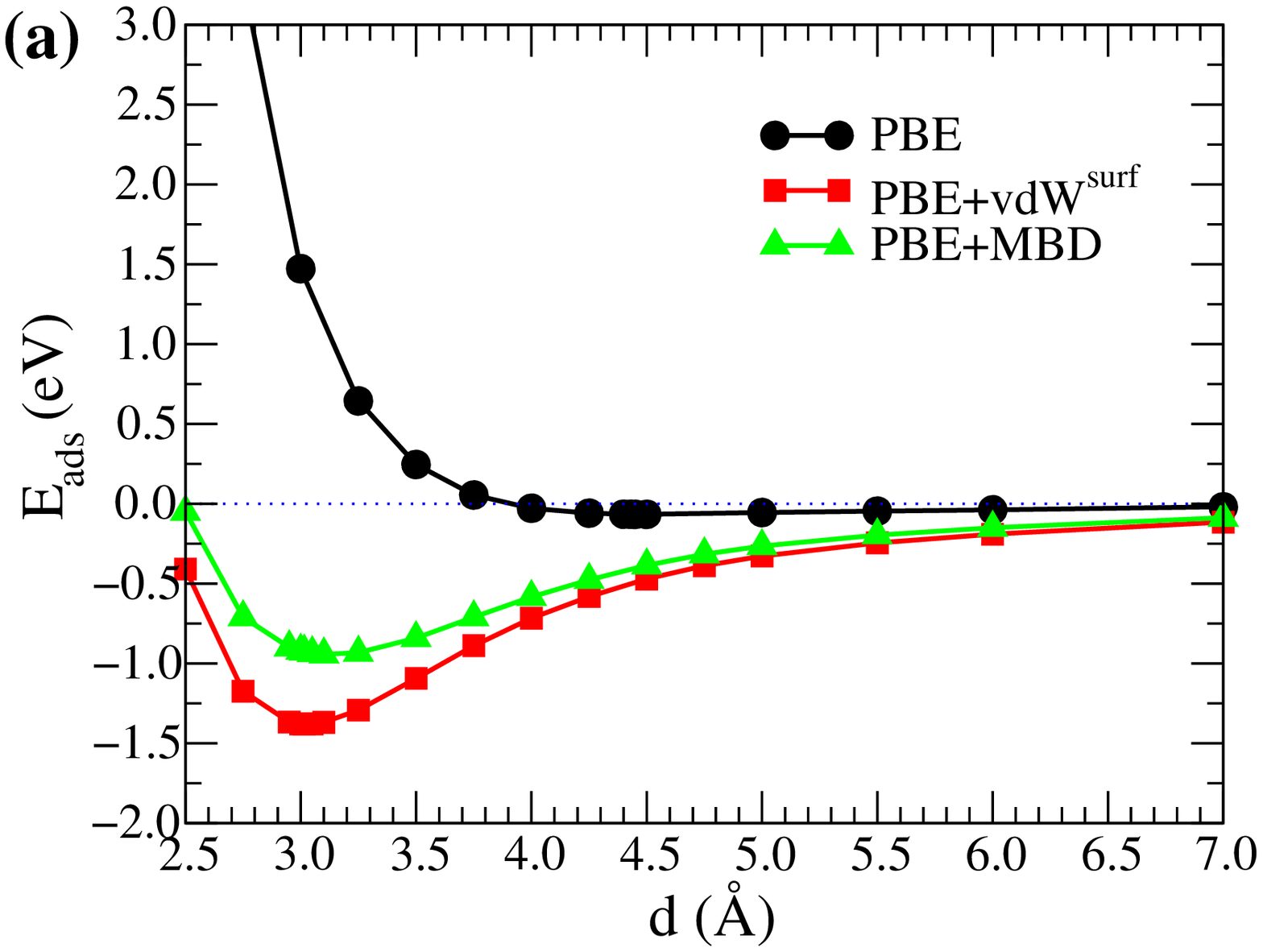} 
\includegraphics[scale=0.41]{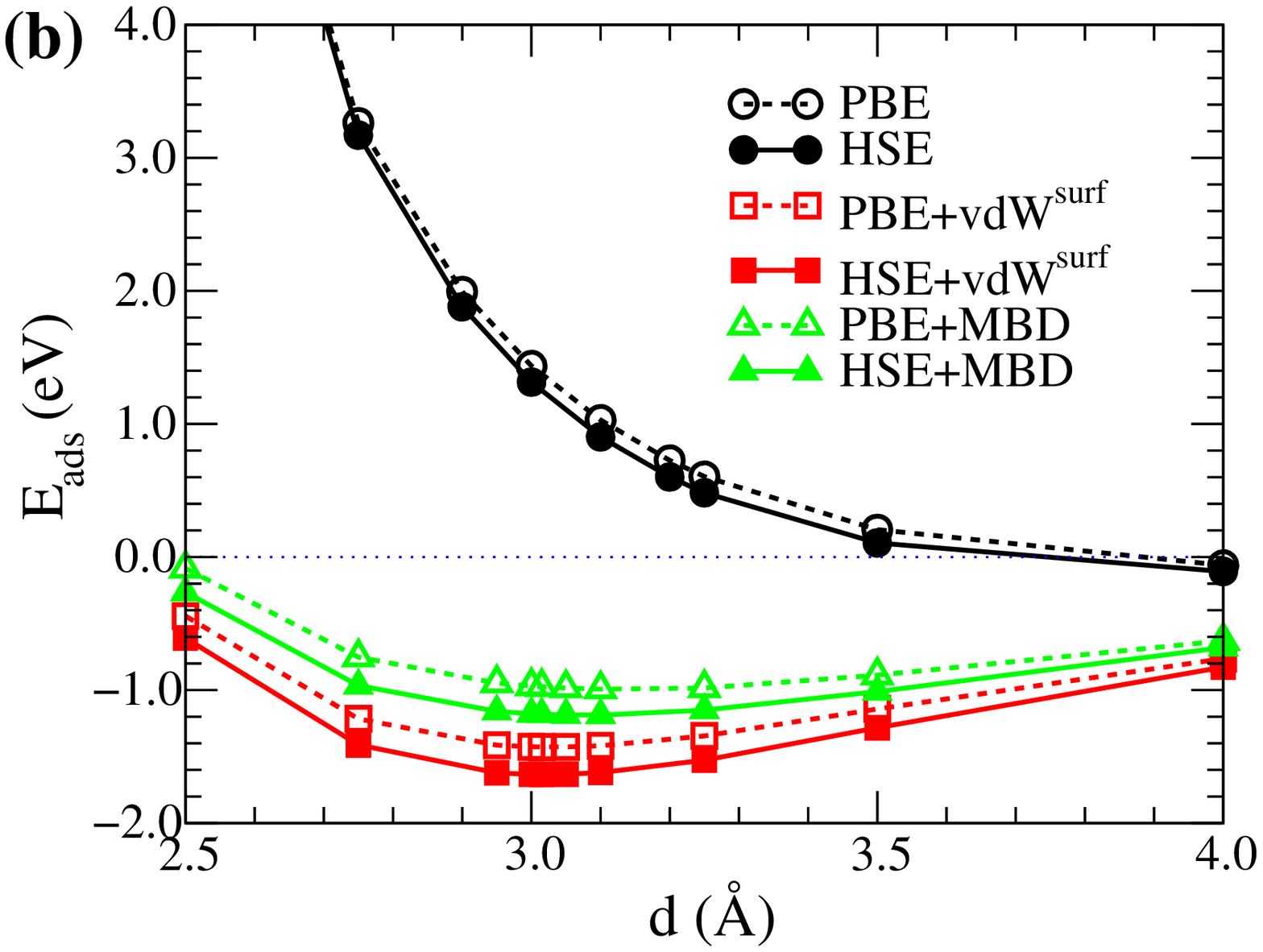} 
\caption{\label{ANT-binding-energy} Adsorption energy $E_{\rm ads}$ of hcp-anthracene/Ag(111) as a function of the adsorption height $d$, 
computed using the (a) PBE, PBE+\vdW and PBE+MBD, and (b) HSE, HSE+\vdW and HSE+MBD approaches. 
The geometry of the system was constrained to the PBE equilibrium geometry in the PBE/HSE calculations and to the 
PBE+\vdW geometry in the calculations with the PBE/HSE+\vdW and PBE/HSE+MBD methods, and only the adsorption height was changed.}
\end{figure}

The failure of PBE to describe the interaction and structural properties of anthracene/Ag(111) 
can also be seen in Fig.~\ref{ANT-binding-energy}(a), which shows the adsorption energy $E_{\rm ads}$ of hcp-anthracene/Ag(111), computed 
using different approaches, as a function of the adsorption height $d$; 
the geometry of the system was constrained to the PBE equilibrium geometry in the PBE calculations and to the 
PBE+\vdW geometry in the calculations with the PBE+\vdW and PBE+MBD methods, and only the adsorption height was changed.  
``Pure'' PBE calculations, without inclusion of vdW corrections, predicted 
repulsive interaction for distances smaller than 4~\AA, and 
very small attractive adsorption energies ($<0.07$~eV) for larger distances, 
suggesting that the molecule 
does not 
bind to the surface. 
The inclusion of vdW corrections, on the other hand, pulled the molecule closer to the surface, 
bringing the adsorption height to values between 3.0 and 3.1~\AA \ (Fig.~\ref{ANT-binding-energy}(a)); 
the curves obtained using the PBE+\vdW and 
PBE+MBD methods exhibit pronounced minima of about $-1.4$ and $-0.9$~eV, respectively.  
To the best of our knowledge, the adsorption height and energy of anthracene/Ag(111) have not yet been determined experimentally. 
Nevertheless, we note that our adsorption energies computed using the PBE+\vdW ($E_{\rm ads}=-1.380$~eV) and PBE+MBD ($E_{\rm ads}=-0.943$~eV) methods 
are within the range of experimental values 
reported 
for benzene/Ag(111), $0.68\pm 0.05$~eV \cite{Liu-PhysRevLett.115.036104, Maurer-ProgSurfSci91.72}, 
naphthalene/Ag(111), $1.03\pm 0.05$~eV \cite{Rockey-SurfSci601-2307, Maurer-ProgSurfSci91.72}, 
and pentacene/Ag(111), $1.5$~eV \cite{Dougherty-JPhysChemC112.20334}. 
The adsorption height obtained using the PBE+\vdW method, 3.015~\AA, 
is also close to that reported for benzene/Ag(111), $3.04\pm 0.02$~\AA \ \cite{Liu-PhysRevLett.115.036104}. 
Previous theoretical studies \cite{Liu-PhysRevLett.115.036104,Maurer-JCP143.102808} 
have shown that the \vdW approach overestimates the adsorption energy of atoms and molecules adsorbed on metal surfaces, 
while the MBD method, which goes beyond the pairwise method and includes many-body dispersion effects, 
predicts results in better agreement with experiments; therefore, we expect that our PBE+MBD calculations would yield more accurate values for the 
adsorption energy and height of realistic anthracene/Ag(111) systems.

We also examined the effects of including part of the short-range exact exchange, by using the HSE06 functional, 
on the adsorption energy of anthracene/Ag(111). As can be seen in Fig.~\ref{ANT-binding-energy}(b), 
the reduction of the self-interaction error leads to an increase of about 0.2~eV in the absolute values of the adsorption energy, with both 
the \vdW and MBD methods. The same behavior was recently observed for benzene/Ag(111) \cite{Liu-PhysRevLett.115.036104}; 
in this case, the adsorption energy obtained with HSE+MBD was in better agreement with experimental results than that obtained with the PBE+MBD method. 
In our case, for anthracene/Ag(111), the adsorption energy obtained using the HSE+MBD method ($E_{\rm ads}$=$-1.18$~eV) is still within range of the experimental values reported 
for other oligoacenes, such as benzene, naphthalene and pentacene, adsorbed on the Ag(111) surface (0.68--1.5~eV) 
\cite{Liu-PhysRevLett.115.036104, Maurer-ProgSurfSci91.72, Rockey-SurfSci601-2307, Maurer-ProgSurfSci91.72, Dougherty-JPhysChemC112.20334}. 

\begin{figure}[!h]
\centering
\includegraphics[scale=0.4]{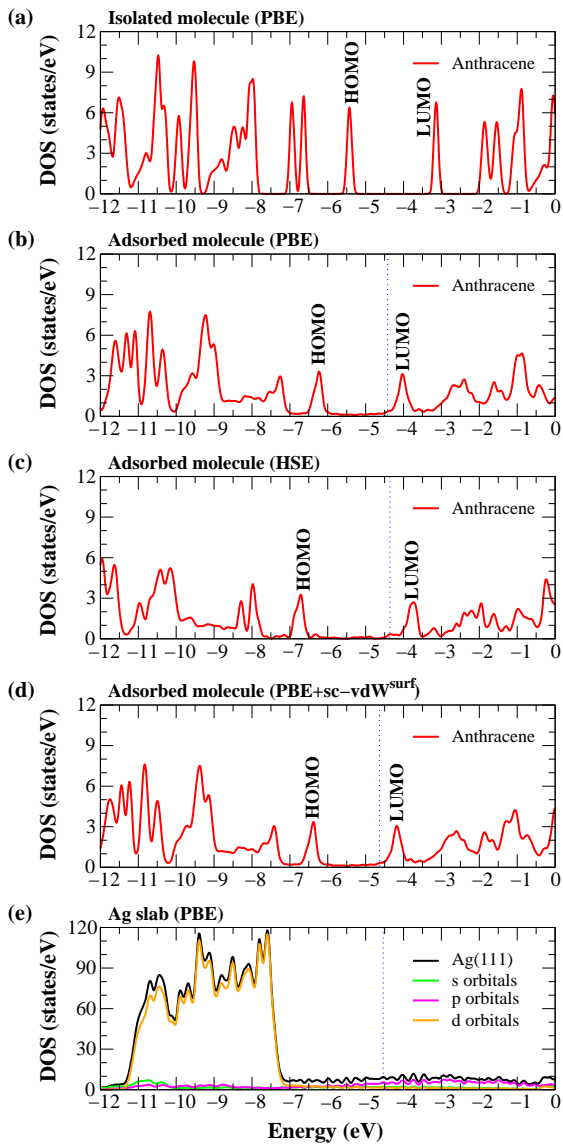} 
\caption{\label{ANT-dos} Calculated density of states (DOS) of the anthracene/Ag(111) system in the optimized PBE+\vdW geometry. 
(a)~Isolated anthracene, (b-d)~density of states projected onto the molecular orbitals of anthracene/Ag(111) obtained with (b)~PBE and (c)~HSE functionals and (d)~the PBE+sc-\vdW method, 
and (e)~clean Ag(111) surface. 
The zero on the energy scale corresponds to the vacuum level. The dashed lines indicate the position of the Fermi level.} 
\end{figure}

In addition to the structural properties, we also investigated the electronic properties of anthracene/Ag(111). 
Figure~\ref{ANT-dos} shows the density of states (DOS) of the isolated anthracene (Fig.~\ref{ANT-dos}(a)) 
and the adsorbed molecule in the PBE+\vdW optimized geometry (Figs.~\ref{ANT-dos}(b)-(d)); 
for comparison, the DOS of the adsorbed molecule in the PBE geometry is shown in Fig.~S1 of the Supplemental Material. 
For the adsorbed system in the PBE+\vdW geometry we present the DOS obtained 
without including vdW interactions (``pure'' PBE and HSE calculations, shown in Figs.~\ref{ANT-dos}(b) and (c), respectively) 
and with vdW corrections using the self-consistent \vdW (sc-\vdW) method \cite{Ferri-PhysRevLett.114.176802} (Fig.~\ref{ANT-dos}(d)).  
The highest occupied (HOMO) and lowest unoccupied (LUMO) orbitals are formed by bonding 
$\pi$ and antibonding $\pi^*$ superpositions of the carbon $p_z$ orbitals above and below 
the molecular plane. 
Comparing the DOS of the isolated (Fig.~\ref{ANT-dos}(a)) and adsorbed (Fig.~\ref{ANT-dos}(b)) anthracene, 
we found that the presence of the substrate induced only a small decrease in the HOMO-LUMO gap ($\sim$0.13~eV) and 
a slight broadening in the HOMO and LUMO peaks (see also Fig.~S2 of the Supplemental Material). 
On the other hand, the states located between 2.5 and 7~eV below the Fermi level are strongly affected the hybridization between the molecular orbitals 
(mainly composed of C $p$ orbitals (Fig.~S2)) and the Ag $d$ states, which are the dominant states between 2.5 and 7~eV (Fig.~\ref{ANT-dos}(e)). 
Figs.~\ref{ANT-dos}(b) and (c) also show that both PBE and HSE calculations predicted that the LUMO remains unoccupied on the Ag(111) surface, 
indicating that there is no charge transfer between the substrate and the molecule and that anthracene is physisorbed on Ag(111), which is consistent with 
previous experimental studies \cite{Yannoulis-SurfSci241.325, Frank-JCP89.7569}.

As expected, the PBE functional significantly underestimates the HOMO-LUMO band gap of the molecule; 
for the isolated molecule (Fig.~\ref{ANT-dos}(a))  we found 
a HOMO-LUMO gap of 2.31~eV, in good agreement with previous theoretical result (2.25~eV) \cite{Blase-PhysRevB.83.115103} 
but $\sim$66\% underestimated with respect to the experimental value (6.9~eV) \cite{Blase-PhysRevB.83.115103}. 
The HSE hybrid functional brings the HOMO to lower energies, yielding a larger band gap, 2.97~eV (still $\sim$57\% smaller than the experimental value). 
HSE also improves the prediction of the HOMO position of the adsorbed molecule; 
while with the PBE functional we obtained HOMO located at 1.81~eV below the Fermi level (Fig.~\ref{ANT-dos}(b)), 
our HSE DOS (Fig.~\ref{ANT-dos}(c)) shows a pronounced peak at 2.33 eV below the Fermi level, which is in good agreement with 
the experimental observation of a $\pi$-character band at the top of the valence band located at 2.5 eV below the Fermi edge \cite{Yannoulis-SurfSci241.325}. 
Both PBE and HSE functionals, however, fail to describe the position of the LUMO; we found that the LUMO is located right above the 
Fermi level (0.37~eV from PBE and 0.64~eV from HSE calculation) whereas inverse photoemission measurements \cite{Yannoulis-SurfSci241.325} 
have shown that the LUMO is located between 2 and 3~eV 
above the Fermi level. This underestimation is mainly due to the inability of Kohn-Sham energy levels to describe excited states.  

We examined the effect of including vdW corrections in the description of the electronic structure of anthracene/Ag(111), 
after the correct geometry has been obtained. 
Thus, considering the anthracene/Ag(111) system at the optimized PBE+\vdW geometry, we compare 
the DOS of the adsorbed molecule obtained using the PBE functional without vdW corrections (Fig.~\ref{ANT-dos}(b)) 
with that obtained using the PBE+sc-\vdW method (Fig.~\ref{ANT-dos}(d)); we found that the inclusion of vdW corrections 
has, apart from a shift of the Fermi level, no significant effect on the 
electronic properties of the system (see also Fig.~S3 of the Supplemental Material). 
Our results therefore show that, although a proper treatment of the vdW interactions is crucial 
to correctly describe the stability and structural properties of anthracene/Ag(111), 
vdW interactions have little effects on the electronic structure of the system once the correct geometry has been obtained. 
The effect of the vdW interactions on the electronic structure is therefore indirect, 
via the correct description of the geometry of the system; 
the DOS of the adsorbed molecule at the PBE geometry (see Fig. S1(a))   
is similar to the DOS of the isolated molecule (Fig.~\ref{ANT-dos}(a)), while the DOS of the adsorbed molecule at the PBE+\vdW geometry (Fig.~\ref{ANT-dos}(b)) is broader 
with the LUMO state located closer to the Fermi level (see also Fig.~S1 for a comparison between the DOS of the adsorbed molecule at the PBE and PBE+\vdW geometries).  
We note that by electronic structure we mean the relative position of the bands with respect to each other and the Fermi level; 
a shift of the energetic reference point, {\it i.e.} the vacuum energy, 
is much more difficult to detect from a plot of the DOS. 

The redistribution of charge upon adsorption is most sensitively reflected in a change of the 
surface dipole, and hence in the work function. Therefore, we computed the change in the work function of the Ag(111) surface due to the adsorption of an anthracene monolayer. 
All calculations were performed using the PBE+\vdW optimized geometry. 
For the work function of the clean Ag(111) surface we obtained 4.43~eV without including vdW corrections (``pure'' PBE calculations)  
and 4.81~eV with the PBE+sc-\vdW method, 
which are in good agreement with previous theoretical results ($4.44$~\cite{Ferri-PhysRevLett.114.176802} and $4.45$~eV~\cite{Romaner-NJP11.053010} 
with PBE and $4.74$~eV~\cite{Ferri-PhysRevLett.114.176802} with the PBE+sc-\vdW approach) and with 
experimental values ($4.45$--$4.90$~eV \cite{Koch-OrgElectronics7.537, Michaelson-JAP48.4729, Duhm-OrgElect9.111, Zou-SurfSci600-1240, Frank-JCP89.7569}). 
The higher work function in the \vdW calculations indicates that the electronic charge density 
extends somewhat further outside the surface. 
As shown and discussed by Ferri et al. in Ref.~\onlinecite{Ferri-PhysRevLett.114.176802}, the inclusion of long-range correlation effects in the self-consistent electronic structure calculation 
leads to a net accumulation of the charge density 
above the surface and between the metal layers and to a depletion at the top metal layers, compared to a PBE calculation with vdW interactions treated by pairwise potentials. 
This lowers the Fermi level (due to the Coulomb attraction between the metal layers) 
and shifts the vacuum level to higher values (due to the extension of the charge density above the surface), consequently increasing the work function. 
Our calculations show that the adsorption of anthracene on Ag(111) leads to a significant reduction in the work function of the surface; 
using the PBE+sc-\vdW method we found 3.77~eV 
for the work function of anthracene/Ag(111), which means a reduction of 1.04~eV with respect to the clean surface, 
while our ``pure'' PBE calculations yielded a decrease of 0.81~eV (from 4.43 to 3.62~eV). 
The large reduction in the work function is due to the push-back effect arising from the Pauli repulsion
between the electron density of the molecule and the surface electrons, which is dominant in physisorbed systems. 
We notice that our PBE and PBE+sc-\vdW results for the change in the work function ($-0.81$ and $-1.04$~eV, respectively) 
are significantly larger than experimental values obtained by measuring the onset current into the sample, 
$-0.5\pm 0.1$~eV~\cite{Yannoulis-SurfSci241.325}, 
and by determining the photoemission kinetic energy width, $-0.68\pm 0.02$~eV~\cite{Gaffney-JPhysChemB105.9031}. 
It should be mentioned, however, that the experimental data available were obtained for multilayer anthracene adsorption and/or for lower 
molecular density. 
We also estimated the change in the positions of the anthracene HOMO and LUMO states 
with respect to the vacuum level upon adsorption: 
from PBE calculations we obtained shifts of 0.80 and 0.92~eV towards lower energies for HOMO and LUMO, respectively, 
while PBE+sc-\vdW yielded shifts of $0.92$ and $1.06$~eV. Previous angle-resolved photoemission and inverse photoemission measurements 
predicted rigid shifts to lower energies of about 0.9~eV for the occupied states and 1.1~eV for the unoccupied orbitals \cite{Yannoulis-SurfSci241.325}.

\subsection{Pentacene on Ag(111)}

\begin{figure}[!t]
\centering
\includegraphics[scale=0.16]{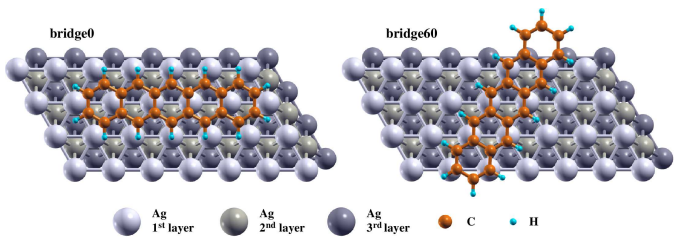}
\caption{\label{PEN-topview} Ball and stick representation of the bridge0 and bridge60 configurations for pentacene/Ag(111).}
\end{figure}

Pentacene monolayers have been experimentally observed to adsorb parallel to the Ag(111) surface \cite{Duhm-ApplMatInt5.9377, Lu-JPCM28.094005, Eremtchenko-PhysRevB.72.115430}; 
therefore, in this work we considered only planar adsorption for pentacene/Ag(111). 
We examined eight adsorption sites: 
four configurations (bridge0, top0, hcp0, and fcc0) in which the molecular long axis is parallel to the larger vector of the surface unit cell 
and four configurations (bridge60, top60, hcp60, and fcc60) in which the molecular long axis is rotate $60^{\circ}$ with respect to the former configurations. 
Figure~\ref{PEN-topview} shows a schematic top view of the configurations bridge0 and bridge60. 

\begin{figure}[!b]
\centering
\includegraphics[scale=0.19]{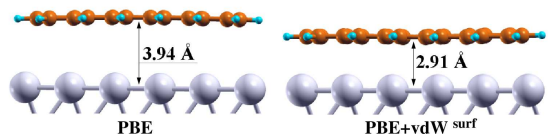}\\
\caption{\label{PEN-sideview} Equilibrium geometries of pentacene/Ag(111) (bridge60 configuration) obtained using the PBE (left)
and PBE+vdW$^{\rm surf}$ (right) methods.}
\end{figure}

We optimized all configurations using the PBE and PBE+\vdW methods and we found, with both methods, that bridge60 is the most favorable configuration (see Table~\ref{table-PEN-conf}). 
The differences in total energy between bridge60 and the other configurations obtained using the PBE+\vdW method (15--226~meV) are significantly larger than those 
obtained in the ``pure'' PBE calculations (1--24~meV). 
This is mainly due to the smaller adsorption heights predicted by the PBE+\vdW method: as can be seen in Table~\ref{table-PEN-conf}, 
PBE+\vdW method yielded adsorption heights between 2.8 and 3.0~\AA, while the PBE values are larger than 3.9~\AA. 
The small difference in total energy between the configurations 
bridge60, hcp60, and hcp0 ($\sim$15~meV with the PBE+\vdW method) also suggests that different adsorption configurations 
may coexist in pentacene/Ag(111), as observed in experiments \cite{Dougherty-JPhysChemC112.20334}. 
Both PBE and PBE+\vdW, however, predict flat conformation for pentacene/Ag(111), as can be seen in Fig.~\ref{PEN-sideview}, in agreement with experimental 
observations \cite{Duhm-ApplMatInt5.9377, Lu-JPCM28.094005, Eremtchenko-PhysRevB.72.115430}. 
Our results are also in good agreement with previous theoretical studies: 
from PBE calculations we found $d=3.938$~\AA \ and $E_{\rm ads}=-0.119$~eV, which compare well with previous GGA results ($d=3.7$--$4.12$~\AA \ and 
$E_{\rm ads}$ within the range of $-0.078$ to $-0.108$~eV) \cite{Toyoda-JCP132.134703, Mete-JPhysChemC114.2724}; 
using the PBE+\vdW method we found $d=2.910$~\AA \ and $E_{\rm ads}=-2.396$~eV, which is in good agreement with previous results obtained by Toyoda et al. 
\cite{Toyoda-JCP132.134703} using the pairwise DFT-D method ($d=2.9$~\AA \ and $E_{\rm ads}=-2.28$~eV), 
but differ from the values obtained by the same authors 
using the nonlocal vdW-DF method ($d=3.7$~\AA \ and $E_{\rm ads}=-1.62$~eV) \cite{Toyoda-JCP132.134703}. 
It should be pointed out that the vdW-DF method tends to overestimate 
the adsorption heights, even though it provides reliable adsorption energies \cite{Li-PhysRevB.85.121409, Maurer-JCP143.102808}---recent vdW-DF functionals 
(e.g., optB86b-vdW and rev-vdW-DF2), however, have been shown to give both accurate adsorption height and energy~\cite{Bjork-ChemPhysChem15.1439}. 
As can be seen in Table~\ref{table-PEN-conf}, 
the inclusion of many-body effects leads to a reduction of the adsorption energy to $E_{\rm ads}=-1.652$~eV, 
in close agreement to the vdW-DF value. 

\begin{table}[!h]
\centering
\caption{\label{table-PEN-conf}Relative total energies $\Delta E_T$ (in eV), 
adsorption energies $E_{\rm ads}$ (in eV), and adsorption heights $d$ (in \AA)  
for pentacene/Ag(111) at different adsorption sites, computed using the PBE and PBE+\vdW methods. 
The equilibrium orientation of the molecule with respect to the substrate (flat or tilted) is indicated within brackets. 
$\Delta E_T$ is calculated as the difference in total energy of a given configuration and the most stable configuration (bridge60).  
The adsorption height is taken as the vertical distance between the center of mass of the molecule 
and the 
average positions of the Ag atoms in the topmost layer. 
The values in parentheses in column 6 are the adsorption energies obtained with the PBE+MBD method at the PBE+\vdW geometry.}
\begin{ruledtabular}
\begin{tabular}{lcccccc}
     & \multicolumn{3}{c}{PBE} & \multicolumn{3}{c}{PBE+\vdW}\\
\cline{2-4}\cline{5-7}
site & $\Delta E_T$ & $E_{\rm ads}$ & $d$ & $\Delta E_T$  & $E_{\rm ads}$ & $d$\\
\hline
bridge0 & $0.017$ & $-0.102$ & $4.097$    & $0.174$ & $-2.189$   & $2.956$ \\
       &       &	& [flat]   &       & ($-1.400$) & [flat] \\
top0    & $0.017$ & $-0.102$ & $4.081$    & $0.226$ & $-2.119$   & $2.965$ \\
       &       &        & [flat]   &       & ($-1.295$) & [flat] \\
hcp0    & $0.002$ & $-0.117$ & $3.942$    & $0.015$ & $-2.378$   & $2.902$ \\
       &       &        & [flat]   &       & ($-1.646$) & [flat]\\
fcc0    & $0.001$ & $-0.118$ & $3.943$    & $0.052$ & $-2.331$   & $2.823$ \\
       &       &        & [flat]   &       & ($-1.585$) & [flat]\\
bridge60 & $0.000$ & $-0.119$ & $3.938$    & $0.000$ & $-2.396$   & $2.910$ \\
       &       &	& [flat]   &       & ($-1.652$) & [flat] \\
top60    & $0.024$ & $-0.095$ & $3.946$    & $0.226$ & $-2.120$   & $2.964$ \\
       &       &        & [flat]   &       & ($-1.295$) & [flat] \\
hcp60    & $0.011$ & $-0.111$ & $3.947$    & $0.015$ & $-2.379$   & $2.901$ \\
       &       &        & [flat]   &       & ($-1.646$) & [flat]\\
fcc60    & $0.006$ & $-0.113$ & $3.943$    & $0.034$ & $-2.336$   & $2.921$ \\
       &       &        & [flat]   &       & ($-1.600$) & [flat]\\
\end{tabular}
\end{ruledtabular}
\end{table}

\begin{figure}[!h]
\centering
\includegraphics[scale=0.41]{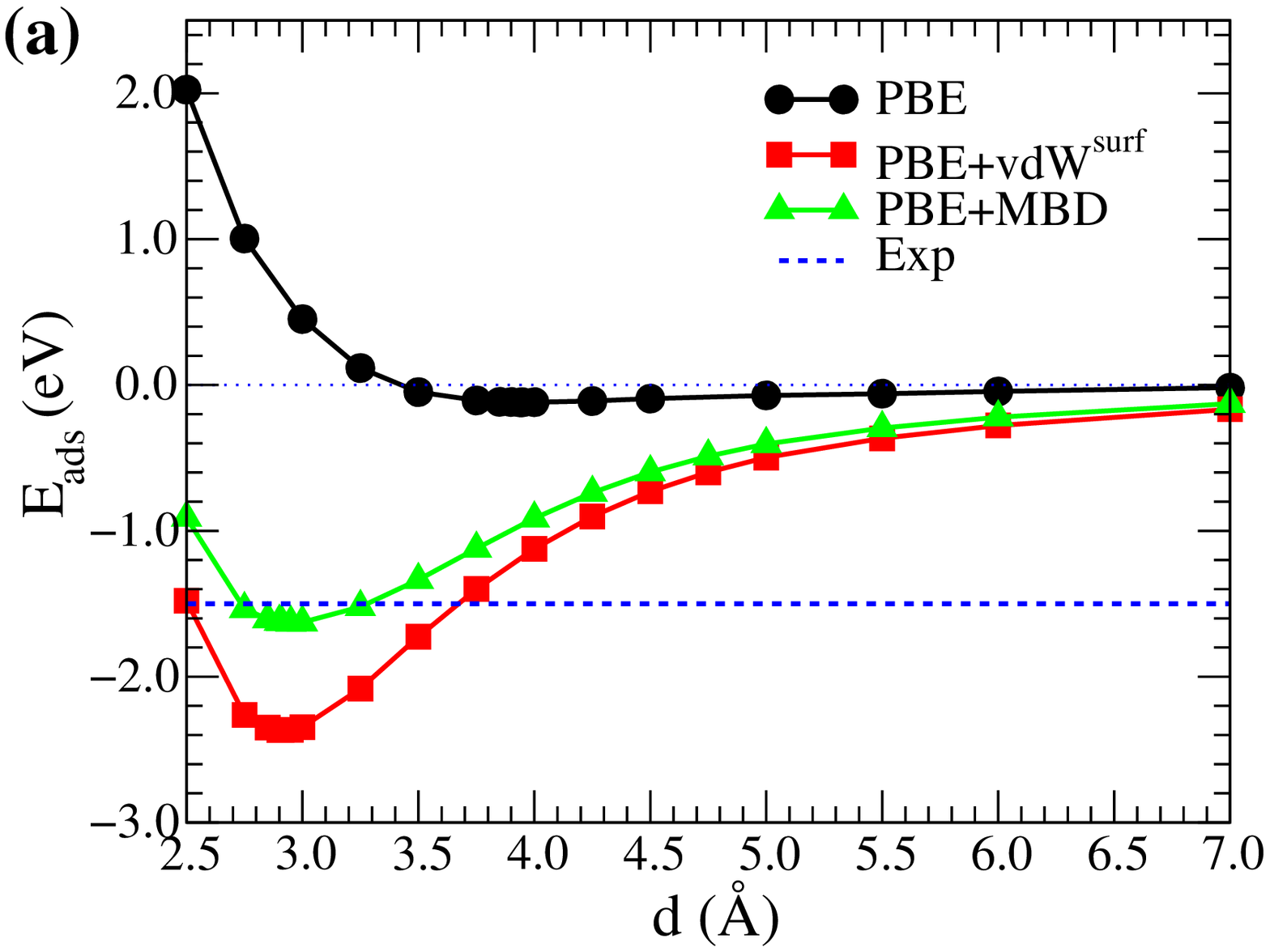} 
\includegraphics[scale=0.41]{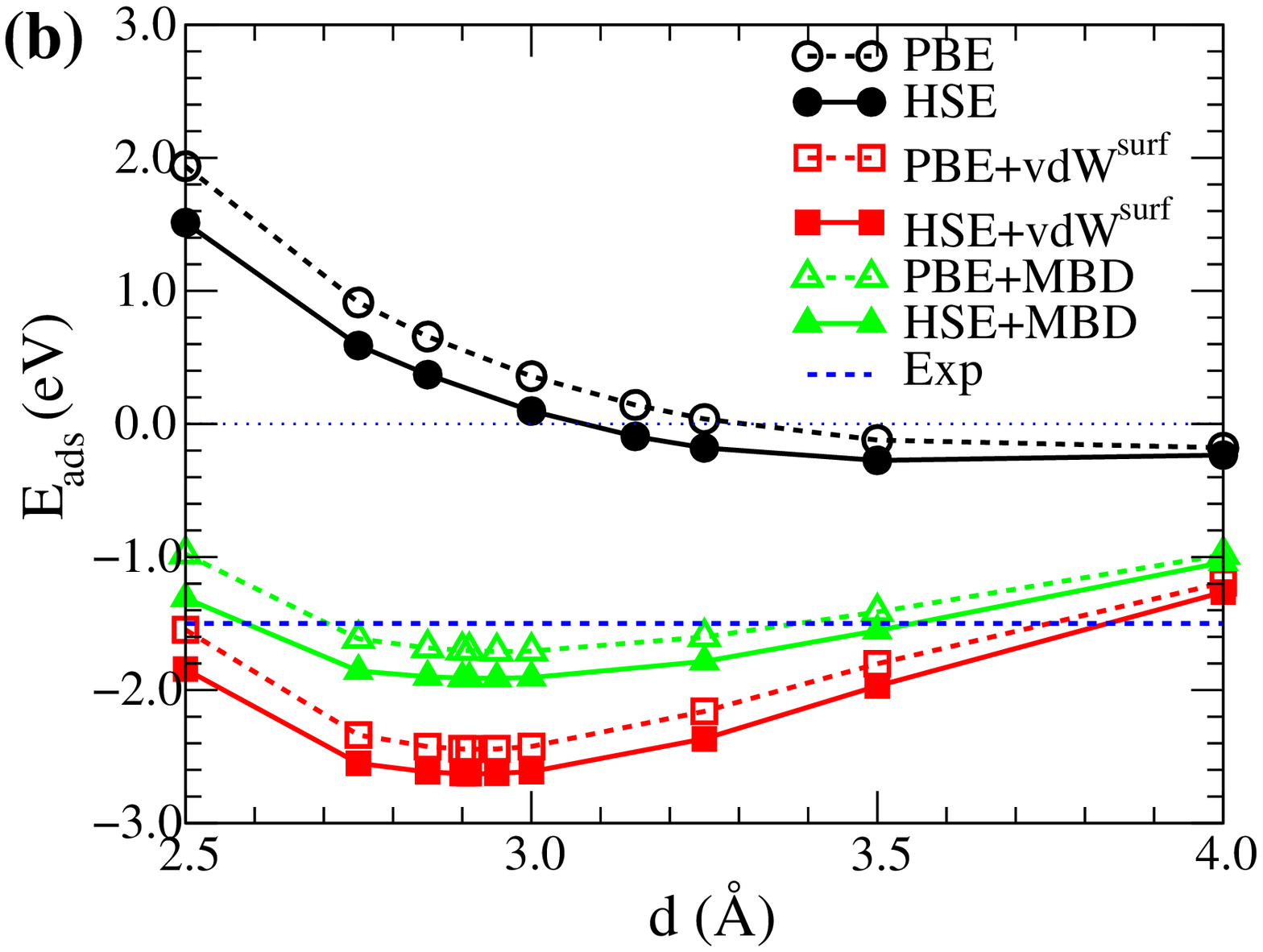} 
\caption{\label{PEN-binding-energy} Adsorption energy $E_{\rm ads}$ of bridge60-pentacene/Ag(111) as a function of the adsorption height $d$, 
computed using the (a) PBE, PBE+\vdW and PBE+MBD, and (b) HSE, HSE+\vdW and HSE+MBD approaches. 
The geometry was constrained to the PBE equilibrium geometry in the PBE/HSE calculations and to the 
PBE+\vdW geometry in the calculations with the PBE/HSE+\vdW and PBE/HSE+MBD methods, and only the adsorption height was changed. 
The blue dashed lines represent the experimental value of the desorption energy of pentacene/Ag(111) reported in Ref. \onlinecite{Dougherty-JPhysChemC112.20334}.}
\end{figure}

Figure~\ref{PEN-binding-energy}(a) shows the adsorption energy of bridge60-pentacene/Ag(111) as a function of the adsorption height, computed using different 
methods. Similar to what was observed for anthracene/Ag(111), 
``pure'' PBE calculations predicted low adsorption energies ($<0.12$~eV) and large adsorption heights ($\sim$3.9~\AA) suggesting that the molecule does not bind to the surface. 
The inclusion of the vdW interactions, again, brought the molecule closer to the surface ($\sim$2.9~\AA) and increased the 
absolute value of the adsorption energy. 
Using the PBE+MBD method we found $E_{\rm ads}=-1.652$~eV, in good agreement with both the theoretical result ($-1.62$~eV) obtained by Toyoda et al. 
\cite{Toyoda-JCP132.134703}, who used the vdW-DF method, and with the experimental value of the desorption energy of pentacene/Ag(111), 1.5~eV~\cite{Dougherty-JPhysChemC112.20334}. 
The comparison between our computed adsorption energies and the experimental value of the desorption energy 
shows that (i) the PBE+\vdW method overestimates the adsorption energy, as has been observed for other 
systems \cite{Liu-PhysRevLett.115.036104,Maurer-JCP143.102808}, and (ii) the inclusion of many-body effects is important to correctly describe the interaction between pentacene and 
the Ag(111) surface. Figure~\ref{PEN-binding-energy}(b) shows that the hybrid HSE functional leads to an increase of $\sim$0.2~eV in the absolute values of the adsorption energy, 
similar to what we found for anthracene/Ag(111).  

\begin{figure}[!h]
\centering
\includegraphics[scale=0.4]{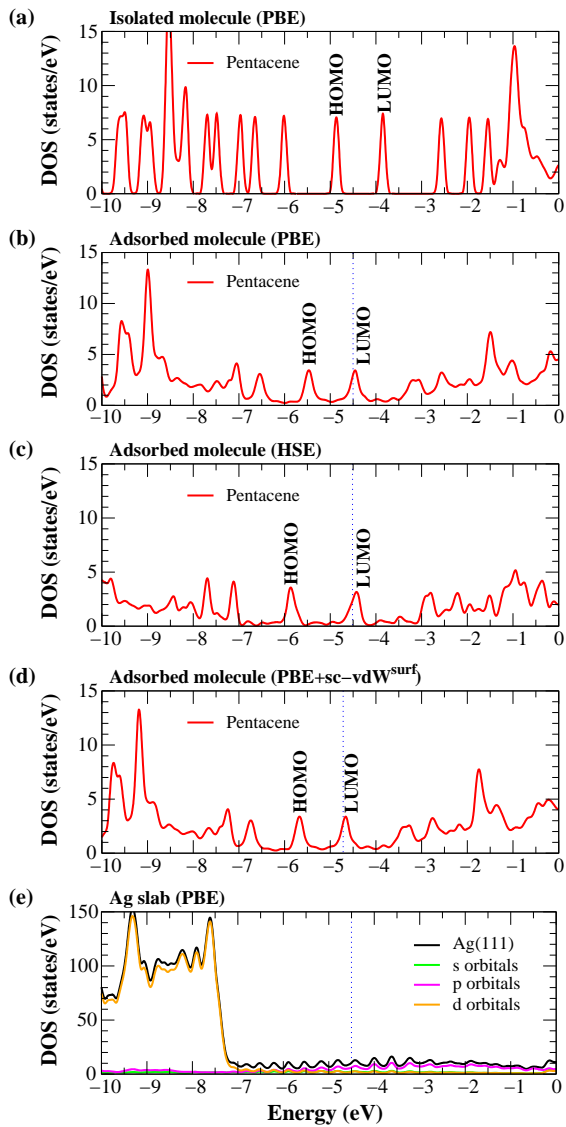} 
\caption{\label{PEN-dos} Calculated density of states (DOS) of the pentacene/Ag(111) system in the optimized PBE+\vdW geometry. 
(a) Isolated pentacene, (b-d) density of states projected onto the molecular orbitals of pentacene/Ag(111) obtained with (b) PBE and (c) HSE functionals and (d)~the PBE+sc-\vdW method, 
and (e) clean Ag(111) surface. 
The zero on the energy scale corresponds to the vacuum level. The dashed lines indicate the position of the Fermi level.} 
\end{figure}

The density of states of bridge60-pentacene/Ag(111) is displayed in Fig.~\ref{PEN-dos}. 
Since the $\pi$ and $\pi^*$ orbitals are more delocalized in pentacene compared to anthracene, 
the energy splitting between HOMO and LUMO is smaller. 
Similar to what was observed for anthracene, the presence of the substrate 
has small effects on the 
relative position of the HOMO and LUMO states (see Figs.~\ref{PEN-dos}(a) and (b) and Fig.~S5 of the Supplemental Material). 
We found, however, that the LUMO level in pentacene/Ag(111) is located closer to the Fermi level than in 
anthracene/Ag(111). Our PBE (Fig.~\ref{PEN-dos}(b)) and HSE (Fig.~\ref{PEN-dos}(c)) calculations predicted that the peak of the LUMO is at 
$\sim$0.06 and 0.07~eV above the Fermi level, respectively. 
Although we expect the LUMO position be underestimated since Kohn-Sham levels are unsuitable to describe excited states, 
this finding suggests that the interaction between molecule and substrate is stronger in pentacene/Ag(111). 
As we will discuss later, this will be reflected in the change of the work function of Ag(111) upon adsorption of pentacene and anthracene; 
pentacene induces a smaller reduction in the work function than anthracene.  
Regarding the occupied states, the HSE functional (Fig.~\ref{PEN-dos}(c)) 
predicted that the HOMO is located at $\sim$1.38~eV below the Fermi level, which is in good agreement with the binding energy of 1.5~eV obtained by 
ultraviolet photoelectron spectroscopy \cite{Koch-OrgElectronics7.537}. 
Our PBE calculations overestimated the position of the HOMO state, placing it at about 
1.0~eV below the Fermi level (Fig.~\ref{PEN-dos}(b)). 

Comparing Figs.~\ref{PEN-dos}(b) and (d) we notice that the inclusion of the vdW interactions using 
the sc-\vdW method does not alter the electronic properties of pentacene/Ag(111) by more than a constant energy shift 
(see also Fig.~S6 of the Supplemental Material). Similar to what was observed for anthracene/Ag(111), 
vdW interactions are important for the description of the stability and geometry of the 
system, but have no significant effect on its electronic structure 
once the correct geometry has been obtained. Again, we found that the effect of the vdW interactions on the electronic structure of pentacene/Ag(111) is indirect, 
via the correct description of the geometry of the system; as can be seen in Fig. S4(a), the DOS of the adsorbed molecule at the PBE geometry is similar to the DOS of 
the isolated molecule (Fig.~\ref{PEN-dos}(a)), whereas the DOS of the adsorbed molecule at the PBE+\vdW geometry (Fig.~\ref{PEN-dos}(b)) 
is signiticantly different, with broader HOMO and LUMO peaks and the LUMO state located closer to the Fermi level 
(see also Fig. S4 for a comparison between the DOS of the adsorbed molecule at the PBE and PBE+\vdW geometries). 
As it will be discussed in the next paragraph, the proximity of the LUMO state 
to the Fermi level strongly affects the change in the work function of the Ag(111) surface due to the adsorption of pentacene molecules.

Lastly, we also computed the change in the work function of the Ag(111) surface due to the adsorption of a pentacene monolayer. 
Using ``pure'' PBE calculations we found that the work function decreases 
from 4.43~eV to 3.98~eV, a reduction of 0.45~eV; with the PBE+sc-\vdW method, we found that the work function decreases by 0.73~eV, changing from 4.81 to 4.08~eV. Our results for the work 
function of the pentacene/Ag(111) system ($\phi=3.98$ and 4.08~eV), as well as the computed reduction ($\Delta \phi=-0.45$ and $-0.73$~eV) 
are in close agreement with experimental values 
reported in Ref.~\onlinecite{Koch-OrgElectronics7.537}, $\phi=3.95$~eV and $\Delta \phi=-0.5$~eV, and in Ref.~\onlinecite{Lu-JPCM28.094005}, 
$\phi=4.00$~eV and $\Delta \phi=-0.59$~eV. 
We noticed, in addition, that the change in the work function of Ag(111) induced by the adsorption of the pentacene monolayer is 
substantially smaller than that caused by the adsorption of anthracene. 
In order to explain this finding, one can think of the work function change as being composed of two opposing components: the first one is the push-back effect, which dominates in anthracene/Ag(111) 
and is present in pentacene/Ag(111) as well; the second contribution originates from an accumulation of charge on the molecule and/or in the space between the molecule and the surface, which tends to 
increase the work function and thus counteracts the push-back effect. 
This latter effect, while being absent for anthracene, starts to play some role for pentacene and larger oligoacenes. 
On the level of our DFT calculations this can be concluded from the hybridization with the substrate states and energetic broadening of the LUMO seen in Fig.~\ref{PEN-dos}. 
Even if the very low energetic position of the LUMO may be an artefact of the DFT calculations, 
we think that the trends observed for the change in work function will persist in a higher-level treatment of the electronic structure:  
also in post-Hartree-Fock methods, we expect induced fractional charges on the adsorbed molecule to counteract the original work function lowering due to the push-back effect. 
The size of possibly induced charges scales with the polarisability of the free molecule. The larger the oligoacenes, the more polarisable they are, 
hence the trend to smaller work function change when larger molecules are adsorbed.

\section{CONCLUSIONS} \label{conclusions}

In summary, we investigated the effects of the vdW interactions on the structural and electronic properties of anthracene/Ag(111) and 
pentacene/Ag(111). We employed two methods to treat the vdW interactions in our calculations: the \vdW approach and the MBD method. 
Our results show that the inclusion of vdW corrections is crucial to correctly describe 
the binding geometries of these systems. For anthracene/Ag(111), in particular, 
``pure'' PBE calculations favored a tilted configuration 
whereas the PBE+\vdW approach yielded a flat-lying geometry, in agreement with previous experimental studies. 
We found, additionally, that vdW interactions strongly affect the adsorption energies and heights of anthracene/Ag(111) and 
pentacene/Ag(111); both PBE+\vdW and PBE+MBD methods predicted adsorption heights considerably smaller (between 2.9 and 3.1~\AA, while PBE results 
are larger than 3.9~\AA) and adsorption energies significantly larger (more than one order of magnitude larger) than the PBE results, 
in better agreement with experimental data and more consistent with previous studies on other oligoacenes adsorbed on the Ag(111) surface. 
We also examined the effect of the vdW interactions on the electronic properties of anthracene/Ag(111) and pentacene/Ag(111), by using the 
self-consistent \vdW method. We found that the inclusion of the self-consistency of the vdW energy in our calculations 
has only tiny effects on the electronic structure of these systems, indicating that the major influence of the vdW interactions is in the stability and 
structural properties of organic/metal systems. 

Analyses of the density of states of anthracene/Ag(111) and pentacene/Ag(111) obtained using the PBE, HSE and PBE+sc-\vdW methods 
revealed that the LUMO level of an anthracene monolayer remains unoccupied upon adsorption on Ag(111)
indicating physisorption for anthracene/Ag(111); in pentacene/Ag(111) the LUMO peak is located right above the Fermi level, suggesting 
stronger interaction between molecule and substrate.  
The stronger molecule-metal interaction in pentacene/Ag(111) explains the smaller reduction in the work function 
of the Ag(111) surface obtained upon pentacene adsorption (PBE: $-0.45$~eV, PBE+sc-\vdW: $-0.73$~eV) when 
compared to that computed for the adsorption of an anthracene monolayer (PBE: $-0.81$~eV, PBE+sc-\vdW: $-1.04$~eV).

\section*{SUPPLEMENTARY MATERIAL}

See supplementary material for further details on the density of states of anthracene/Ag(111) and pentacene/Ag(111). 

\section*{Acknowledgments}

We acknowledge fruitful discussions within the collaborative research center SFB 1242 
``Non-equilibrium dynamics in condensed matter in the time domain'' funded by the 
Deutsche Forschungsgemeinschaft. 
The authors gratefully acknowledge the computing time granted by the Center for 
Computational Sciences and Simulation (CCSS) of the University of Duisburg-Essen and provided on the supercomputer magnitUDE 
(DFG grant INST 20876/209-1 FUGG) at the Zentrum f\"ur Informations- und Mediendienste
(ZIM).

\end{document}